\documentclass{article}
\usepackage{geometry}
\usepackage{hyperref}

\def\bd{\begin{displaymath}}\def\ed{\end{displaymath}}
\def\beq{\begin{equation}}\def\eeq{\end{equation}}
\def\bea{\begin{eqnarray}}\def\eea{\end{eqnarray}}
\def\ba{\begin{array}}\def\ea{\end{array}}
\def\bs{\begin{split}}\def\es{\end{split}}
\def\nn{\nonumber}\def\lb{\label}

\def\al{\alpha}\def\be{\beta}
\def\ph{\phi}
\def\la{\lambda}\def\om{\omega}
\def\ps{\psi}

\def\De{\Delta}
\def\Om{\Omega}

\def\deh{\partial}
\def\ha{{1\over 2}}\def\mo{{-1}}
\def\coo{coordinates }\def\rep{representation }
\def\poi{Poincar\'e }\def\schr{Schr\"odinger }

\def\bx{\bar x}\def\bp{\bar p}\def\hx{\hat x}\def\hp{\hat p}

\def\hpp{\hat \pi}

\def\ubp{(1+\be^2p^2)}\def\ubz{1-\be^2z^2}\def\embc{\sqrt{1-\be^2p^2}}
\def\ubq{\left(1+{\be^2\over4}p^2\right)}
\begin{document}

\

\begin{center}

\LARGE{{Diffeomorphisms in momentum space:  physical implications of different choices of momentum coordinates in the Galilean Snyder model
}} \\
\end{center}

\

\begin{center}

{\sc Giulia Gubitosi$^{1,2}$,  Salvatore Mignemi$^{3,4}$}

\medskip

{$^{1}$ Dipartimento di Fisica ``Ettore Pancini'',
Universit\`a di Napoli Federico II, Napoli, Italy}

{$^{2}$ INFN, Sezione di Napoli}

{$^{3}$ Dipartimento di Matematica, Universit\`a di Cagliari,
via Ospedale 72, 09124 Cagliari, Italy}

{$^{4}$ INFN, Sezione di Cagliari}

 \medskip

e-mail: { \href{mailto:giulia.gubitosi@unina.it}{giulia.gubitosi@unina.it}, \href{mailto:smignemi@unica.it}{smignemi@unica.it}}


\end{center}

\medskip

\begin{abstract}

It has been pointed out that different choices of momenta can be associated to the same noncommutative spacetime model. The question of whether these momentum spaces, related by diffeomorphisms, produce the same physical predictions is still debated. In this work, we focus our attention on a few different momentum spaces that can be associated to the Galilean Snyder noncommutative spacetime model and show that they produce different predictions for the energy spectrum of the harmonic oscillator.
\end{abstract}


  \medskip

\section{Introduction}

The Snyder noncommutative spacetime model was proposed in \cite{Snyder:1946qz} with the aim of removing the divergencies emerging in quantum field theory. The main idea was that spacetime coordinates could acquire a discrete spectrum characterized by a fundamental length scale without breaking  Lorentz invariance. This was the first time noncommuting coordinates were considered in the literature, and this paper anticipated the development of noncommutative geometry by several decades.\footnote{For a short review, see \cite{Mignemi:2020ssi}} While this is well known, it was recently made  explicit \cite{Kowalski-Glikman:2003qjp, Ivetic:2017gte, Ballesteros:2019mxi, Gubitosi:2020znn} that this model was also the first instance were spacetime noncommutativity is linked to momentum space curvature.

In fact, in Snyder's construction the momentum manifold has a de Sitter geometry, and spacetime coordinates are identified with the translation generators over such manifold. Then, spacetime noncommutativity can simply be understood as a consequence of the curvature of the momentum manifold.

Such a feature has recently emerged as a general implication of noncommutative spacetime models \cite{Kowalski-Glikman:2013rxa, Lizzi:2020tci}, including those where relativistic symmetries are described by a (Planck-scale) deformation of the standard Lorentz transformations \cite{Kowalski-Glikman:2002oyi, Gubitosi:2011hgc, Amelino-Camelia:2013sba, Gutierrez-Sagredo:2019ipf, Ballesteros:2021dob}.
These developments converged in the relative locality proposal \cite{Amelino-Camelia:2011lvm, Amelino-Camelia:2011hjg}, which pointed out that locality of interactions becomes an observer-dependent statement in these models, thus   motivating a shift of focus from spacetime to momentum space at least for what concerns  phenomenological applications \cite{Amelino-Camelia:2011uwb, Carmona:2019fwf, Mignemi:2019yzn, Mignemi:2018tdg}.

Because of the prominent role played by momentum space geometry in uncovering the phenomenology of noncommutative spacetime models and of quantum gravity in general, it is important to understand whether diffeomorphisms on momentum space, which induce a change of coordinates on the manifold, have physically relevant implications. When  importing the intuition from general relativity, where spacetime diffeomorphisms  do not modify the physics, one is lead to think that this should be the case also for  momentum space \cite{Banburski:2013jfa, Meljanac:2012pv, Cianfrani:2014fia}. However, while one can indeed identify a duality between features of models with curved momentum space and models with curved spacetime \cite{Amelino-Camelia:2013uya}, the two constructions have fundamental differences, due to the different role played in physics by spacetime and momentum space. For instance, while we know that  the evolution of spacetime geometry is governed by Einstein's equations, at the moment we have no strong motivation to introduce  dynamics on momentum space.  Also, it seems difficult to link momentum diffeomorphisms to a change of observer, contrary to the interpretation of coordinate invariance in general relativity.

In this paper we aim at contributing to the understanding of this issue, by focussing on the phenomenological consequences of different choices of momentum coordinates for the Snyder model.  While the canonical momenta in the Snyder model are given by the Beltrami projective coordinates on the de Sitter manifold, in  recent work \cite{Ballesteros:2019mxi, Gubitosi:2020znn} it was noted that in fact any choice of projective coordinates on the manifold provides viable momentum coordinates, which are commutative and behave classically under the Lorentz transformations. Because the  phase space algebra associated to any such set of coordinates is different, in \cite{Ballesteros:2019mxi, Gubitosi:2020znn} it was suggested that indeed the physical predictions of the Snyder model would depend on this choice, but no explicit example was provided.
The possibility of generalizing Snyder model was already noticed in \cite{Kempf:1994su, Battisti:2010sr} from an algebraic point of view, but its phenomenological were not investigated, except in the context of field theory.

Here, we fill this gap by focussing on what could be considered the simplest nontrivial example, namely that  of a one-dimensional harmonic oscillator.  Because this is a quantum-mechanical setting, we will work in the nonrelativistic limit of the Snyder model.\footnote{It has  been noticed that in a covariant formulation of relativistic quantum mechanics the non-interacting Snyder model is trivial \cite{Amelino-Camelia:2014mea}. The model studied here is interacting and non-relativistic, so we do not expect that the observations of \cite{Amelino-Camelia:2014mea} apply in this case.}
Until recently, this limit was usually taken by simply considering the three-dimensional Euclidean version of the model, thus neglecting the role of the time coordinate \cite{Mignemi:2011gr, Lu:2011it, Ivetic:2015cwa, Leiva:2012az, Quesne:2004pp}. However, in \cite{Ballesteros:2019mxi} it was shown that taking explicitly the limit for $c\to\infty$ gives rise to more complicated relations, where a mixing between time and space variables is still present. The resulting model has been called Galilean Snyder model, to emphasize that it is invariant under  Galilean transformations.
This is reviewed in Section \ref{sec:GalSny}, where we also discuss the different sets of momentum space coordinates that will be used in our analysis. These are the Beltrami projective coordinates, which we mentioned earlier, the Poincar\'e projective coordinates and the embedding coordinates of the de Sitter manifold. In Sections  \ref{sec:Embedding} and \ref{sec:Beltrami} we derive, for each of these three sets of coordinates, the deformed \schr equation for the harmonic oscillator and its eigenvalues. We find that indeed the energy levels differ in the various cases. We comment on our results in the concluding Section \ref{sec:Conclusions}.

In this work we shall limit our investigation to the Snyder model, with positive coupling constant, but similar
calculations can be performed for the anti-Snyder model \cite{Mignemi:2011gr}, taking into account its different properties. We work with a mostly plus metric,  Latin uppercase indices  run from $0$ to $4$,  Greek lowercase indices run from $0$ to $3$ and Latin lowercase indices run from $1$ to $3$.

\section{The Snyder model and its Galilean limit}\label{sec:GalSny}

As we mentioned, Lorentz invariance of the Snyder model is guaranteed by the fact that  spacetime coordinates are identified with the translation generators over a curved manifold with de Sitter geometry. In $3+1$ dimensions, the full algebra of symmetries over such manifold is given in terms of  boosts $K_{i}$, rotations $J_{i}$ and  translations $P_{\alpha}$ as follows:
\begin{equation}
\begin{array}{lll}
[J_i,J_j]= \epsilon_{ijk}J_k ,& \qquad [J_i,P_j]=\epsilon_{ijk}P_k , &\qquad
[J_i,K_j]=\epsilon_{ijk}K_k , \\[4pt]
\displaystyle{
  [K_i,P_0]=P_i  } , &\qquad {[K_i,P_j]=\frac 1{c^2}\, \delta_{ij}  P_0} ,    &\qquad {[K_i,K_j]=-\frac 1{c^2}\,\epsilon_{ijk} J_k} ,
\\[4pt][P_0,P_i]=-\beta^2 \,  K_i , &\qquad   [P_i,P_j]=\beta^2\, \frac 1{c^2}\, \epsilon_{ijk}J_k , &\qquad[P_0,J_i]=0  .
\end{array}
\label{eq:deSitterAlgebra}
\end{equation}
Following \cite{Ballesteros:2019mxi}, we write the speed of light $c$ explicitly in order to be able to easily perform the Galilean limit.  Here $\beta^{2} $ plays the role of curvature of the de Sitter manifold.

The Snyder noncommutative spacetime parametrized by the coordinates $x_\mu$ is then obtained  upon the identification
\begin{equation}
x_0:=- \frac {1} {c}\,P_0,\qquad x_i := - c\,P_i ,\label{eq:SnyderCoordinates}
\end{equation}
so that spacetime coordinates satisfy the following commutation relations induced by the curvature of the de Sitter manifold:
\begin{equation}
\left[x_0,x_i\right]=-\beta^2 \, K_{i} , \qquad \left[x_i,x_j\right]= \beta^2\,\epsilon_{ijk} J_k ,
\end{equation}

In the Snyder model momenta live on the de Sitter manifold. However, the specific choice of coordinates on this manifold to be identified with the physical momenta is not univocal.  Indeed several options, described in detail in \cite{Gubitosi:2020znn},  satisfy the basic properties of being commutative and transforming classically under Lorentz transformations.
The starting point is given by the
 ambient coordinates on the de Sitter manifold,  $\eta_{A}$, that satisfy the constraint
 \begin{equation}
 \eta_4^2-\beta^2\, \eta_0^2+  \frac{\beta^2}{c^2}\left( \eta_1^2+
  \eta_2^2+ \eta_3^2 \right)=1 .
\label{eq:constraint}
\end{equation}
The canonical choice for physical momenta, also used in the original paper by Snyder \cite{Snyder:1946qz}, is given by the (appropriately rescaled) Beltrami projective coordinates, related to the ambient coordinates as follows:
\begin{equation}
p_{0}:= c\,\frac{\eta_0}{\eta_4} ,\qquad p_{i}:= \frac 1 c\, \frac{\eta_i}{\eta_4}  .
\label{eq:Beltrami}
\end{equation}

 With this choice of physical momenta, and turning coordinates and momenta into Hermitian operators, $\hat x_{\mu}\equiv i\hbar x_{\mu}$, $\hat p_{\nu}\equiv p_{\nu}$, that act on the space of functions of momenta $\psi(p)$,  the  Snyder phase space commutation relations read:
 \begin{equation}
\begin{array}{ll}
\left[\hat x_0,\hat x_i\right]=-i\hbar \beta^2 \, \hat K_{i} ,& \qquad \left[\hat x_i,\hat x_j\right]= i\hbar\beta^2\,\epsilon_{ijk} \hat J_k,  \\[2pt]
\displaystyle{ \left[\hat  x_0,\hat p_{\alpha}\right]=i\hbar(-\delta_{0\alpha}+\frac{\beta^2}{c^2} \, \hat p_{0}\hat p_{\alpha}),}& \qquad {   \left[\hat x_i,\hat p_{0}\right]=i\hbar\beta^2 \, \hat p_{0}\hat p_{i},  }
\\[4pt]
 \left[\hat x_i,\hat p_{j}\right]=i\hbar(\delta_{ij}+\beta^2 \, \hat p_{i}\hat p_{j}),   &\qquad  \left[\hat p_{\alpha},\hat p_{\beta}\right]=0,
\end{array}
\label{eq:SnyderBeltrami}
\end{equation}
with   boost and rotation generators given by
\begin{equation}
\hat K_{i}= -\hat x_0 \hat p_{i}+\frac 1 {c^2} \,  \hat x_i \hat p_{0} ,\qquad \hat J_i= \epsilon_{ijk}  \hat x_{j} \hat p_k.\label{LorentzRepBeltrami}
\end{equation}
It is easy to check that this phase space commutators are invariant under Lorentz transformations and that both spacetime coordinates and momenta transform classically under Lorentz symmetries.

In \cite{Gubitosi:2020znn} different choices for the physical momenta were analyzed. In particular, it was noted that any set of projective coordinates for the de Sitter manifold would have the required properties and so be a viable option, all choices being related by a momentum space diffeomorphism. For example, one could define momenta via the Poincar\'e projective coordinates:
\begin{equation}
\tilde p_{0}:= c\,\frac{2 \eta_0}{1+\eta_4} ,\qquad \tilde p_{i}:= \frac 1 c\, \frac{2 \eta_i}{1+\eta_4}  .
\label{eq:Poincare}
\end{equation}
These momenta are related to the Beltrami momenta via the following diffeomorphism:
\begin{equation}
\tilde p_{\alpha}=\frac{2 p_{\alpha}}{1+\sqrt{1-\frac{ \beta^{2}}{c^{2}}\, p_{0}^{2}+ { \beta^{2}} (p_1^2+p_2^2+p_3^2)}},\qquad
 p_{\alpha}=\frac{ \tilde p_{\alpha}}{ {1+\frac{ \beta^{2}}{4 c^{2}}\, \tilde p_{0}^{2}- { \frac{\beta^{2}}{4}} (\tilde p_1^2+\tilde p_2^2+\tilde p_3^2)}}.
\end{equation}
Using as physical momenta the ones related to the Poincar\'e coordinates, the resulting Snyder phase space is
\begin{equation}
\begin{array}{ll}
\left[\hat x_0,\hat x_i\right]= - i\hbar \beta^{2} \,  \hat  K_{i} ,& \qquad \left[\hat x_i,\hat x_j\right]= i\hbar  \beta^{2}\,\epsilon_{ijk}   \hat J_k,  \\[2pt]
\displaystyle{ \left[\hat x_0, {\tilde p}_{\alpha}\right]=i\hbar\left(- \delta_{0\alpha}\biggl( 1+\frac{ \beta^{2}}{4 c^{2}}\,\tilde p_{0}^{2}- \frac{\beta^{2}}{4} (\tilde p_1^2+\tilde p_2^2+\tilde p_3^2)\biggr)-\frac{ \beta^{2}}{2 c^2} \, \tilde p_{0}\tilde p_{\alpha}\right),}& \qquad  \left[\hat x_i,\tilde p_{0}\right]= i\hbar \frac{\beta^{2}}{2} \, \tilde p_{0}\tilde  p_{i},
\\[8pt]
\displaystyle{  \left[\hat x_i,\tilde p_{j}\right]= i\hbar \left(  \delta_{ij}\biggl( 1+\frac{ \beta^{2}}{4 c^{2}}\,\tilde p_{0}^{2}- \frac{\beta^{2}}{4}   (\tilde p_1^2+\tilde p_2^2+\tilde p_3^2 )\biggr)+ \frac{\beta^{2}}{2} \, \tilde p_{i} \tilde p_{j}\right), }  &\qquad  \left[\tilde p_{\alpha},\tilde p_{\beta}\right]=0,
\end{array}
\label{eq:SnyderPoincare}
\end{equation}
where we omitted the hat on the $\tilde p_{\alpha}$ operators in order to simplify the notation. The Lorentz boosts and rotations now take the form
\begin{equation}
\hat K_{i}= \,\frac{-\hat x_0 \tilde p_{i}+\frac 1 {c^2} \,  \hat x_i \tilde p_{0}}{1+\frac{ \beta^{2}}{4 c^{2}}\tilde p_{0}^{2}- \frac{\beta^{2}}{4}  (\tilde p_1^2+\tilde p_2^2+\tilde p_3^2)} \, ,\qquad \hat J_i= \,\frac{\epsilon_{ijk} \hat  x_{j} \tilde p_k}{1+\frac{ \beta^{2}}{4c^{2}}\tilde p_{0}^{2}- \frac{\beta^{2}}{4}  (\tilde p_1^2+\tilde p_2^2+\tilde p_3^2)} \, .
\end{equation}

Finally, one can obtain a quasi-canonical structure for the Snyder phase space by adopting physical momenta that are more directly related to the de Sitter ambient coordinates:
\begin{equation}
\pi_{0}:= c\,\eta_0,\qquad \pi_{i}:= \frac 1 c\,\eta_i  .
\label{eq:ambient}
\end{equation}

In this case, the Snyder phase space algebra reads:
 \begin{equation}
\begin{array}{ll}
\left[\hat x_0,\hat x_i\right]=- i\hbar\beta^{2} \,\hat K_{i} ,& \qquad \left[\hat x_i,\hat x_j\right]=i\hbar \beta^{2}\,\epsilon_{ijk} \hat J_k,  \\[4pt]
\left[\hat x_\alpha,\hat \pi_{\beta}\right]=i\hbar \eta_{\alpha \beta} \sqrt{1+\frac{\beta^{2}}{c^{2}}\hat \pi_{0}^{2}-\beta^{2} (\hat \pi_{1}^{2}+\hat \pi_{2}^{2}+\hat \pi_{3}^{2})},  &\qquad  \left[\hat \pi_{\alpha},\hat \pi_{\beta}\right]=0 ,
\end{array}
\label{eq:SnyderAmbient}
\end{equation}
and the Lorentz generators  are given by
\begin{equation}
\hat K_{i}=  \frac{-\hat x_0 \hat \pi_{i}+\frac 1 {c^2} \,  \hat x_i\hat \pi_{0}}{\sqrt{1+\frac{\beta^{2}}{c^{2}}\hat\pi_{0}^{2}-\beta^{2} (\hat\pi_{1}^{2}+\hat\pi_{2}^{2}+\hat\pi_{3}^{2})}} ,\qquad \hat J_i= \frac{\epsilon_{ijk}  \hat x_{j} \hat\pi_k}{\sqrt{1+\frac{\beta^{2}}{c^{2}}\hat\pi_{0}^{2}-\beta^{2} (\hat\pi_{1}^{2}+\hat\pi_{2}^{2}+\hat\pi_{3}^{2})}} .\label{LorentzRepAmbient}
\end{equation}

Having kept track of the factors of $c$ throughout our brief summary, it is easy to work out the Galilean limit $c\to \infty$ of the Snyder model in either set of momentum coordinates. Using the Beltrami momenta, the Galilean limit gives the following phase space relations:
 \begin{equation}
\begin{array}{ll}
\left[\hat x_0,\hat x_i\right]=-i\hbar \beta^2 \, \hat K_{i} ,& \qquad \left[\hat x_i,\hat x_j\right]= i\hbar\beta^2\,\epsilon_{ijk} \hat J_k,  \\[2pt]
\displaystyle{ \left[\hat  x_0,\hat p_{\alpha}\right]=-i\hbar\delta_{0\alpha},}& \qquad {   \left[\hat x_i,\hat p_{0}\right]=i\hbar\beta^2 \, \hat p_{0}\hat p_{i},  }
\\[4pt]
 \left[\hat x_i,\hat p_{j}\right]=i\hbar(\delta_{ij}+\beta^2 \, \hat p_{i}\hat p_{j}),   &\qquad  \left[\hat p_{\alpha},\hat p_{\beta}\right]=0,
\end{array}
\label{eq:GalileanSnyderBeltrami}
\end{equation}
where now the boost generator is the Galilean one:
\begin{equation}
\hat K_{i}= -\hat x_0 \hat p_{i}\,.\label{GalileiBoostBeltrami}
\end{equation}
Using the momenta given by the Poincar\'e projective coordinates, the phase space algebra in the Galilean limit reads:
\begin{equation}
\begin{array}{ll}
\left[\hat x_0,\hat x_i\right]= - i\hbar \beta^{2} \,  \hat  K_{i} ,& \qquad \left[\hat x_i,\hat x_j\right]= i\hbar  \beta^{2}\,\epsilon_{ijk}   \hat J_k,  \\[2pt]
\displaystyle{ \left[\hat x_0, {\tilde p}_{\alpha}\right]=-i\hbar \,\delta_{0\alpha}\biggl[ 1- \frac{\beta^{2}}{4} (\tilde p_1^2+\tilde p_2^2+\tilde p_3^2)\biggr],}& \qquad  \left[\hat x_i,\tilde p_{0}\right]= i\hbar \frac{\beta^{2}}{2} \, \tilde p_{0}\tilde  p_{i},
\\[8pt]
\displaystyle{  \left[\hat x_i,\tilde p_{j}\right]= i\hbar \left[  \delta_{ij}\biggl( 1- \frac{\beta^{2}}{4}   (\tilde p_1^2+\tilde p_2^2+\tilde p_3^2 )\biggr)+ \frac{\beta^{2}}{2} \, \tilde p_{i} \tilde p_{j}\right], }  &\qquad  \left[\tilde p_{\alpha},\tilde p_{\beta}\right]=0,
\end{array}
\label{eq:GalileanSnyderPoincare}
\end{equation}
with Galilean boost and rotation generators:
\begin{equation}
\hat K_{i}= \,\frac{-\hat x_0 \tilde p_{i}}{1- \frac{\beta^{2}}{4}  (\tilde p_1^2+\tilde p_2^2+\tilde p_3^2)} \, ,\qquad \hat J_i= \,\frac{\epsilon_{ijk} \hat  x_{j} \tilde p_k}{1- \frac{\beta^{2}}{4}  (\tilde p_1^2+\tilde p_2^2+\tilde p_3^2)} \, .
\end{equation}
Finally, using the momenta given by ambient coordinates, the Galilean phase space reads:
 \begin{equation}
\begin{array}{ll}
\left[\hat x_0,\hat x_i\right]=- i\hbar\beta^{2} \,\hat K_{i} ,& \qquad \left[\hat x_i,\hat x_j\right]=i\hbar \beta^{2}\,\epsilon_{ijk} \hat J_k,  \\[4pt]
\left[\hat x_\alpha,\hat \pi_{\beta}\right]=i\hbar\, \eta_{\alpha \beta} \sqrt{1-\beta^{2} (\hat \pi_{1}^{2}+\hat \pi_{2}^{2}+\hat \pi_{3}^{2})},  &\qquad  \left[\hat \pi_{\alpha},\hat \pi_{\beta}\right]=0 ,
\end{array}
\label{eq:SnyderAmbient}
\end{equation}
with Galilean boost and rotation generators
\begin{equation}
\hat K_{i}=  \frac{-\hat x_0 \hat \pi_{i}}{\sqrt{1-\beta^{2} (\hat\pi_{1}^{2}+\hat\pi_{2}^{2}+\hat\pi_{3}^{2})}} ,\qquad \hat J_i= \frac{\epsilon_{ijk}  \hat x_{j} \hat\pi_k}{\sqrt{1-\beta^{2} (\hat\pi_{1}^{2}+\hat\pi_{2}^{2}+\hat\pi_{3}^{2})}} .\label{LorentzRepAmbient}
\end{equation}
As was already emphasized in \cite{Ballesteros:2019mxi}, the Galilean limit does not remove the mixing between space and time components of the phase space, as is instead the case in standard mechanics.
In the following sections we shall study the spectrum of the harmonic oscillator in this limit and compare it with the case of classical quantum mechanics and Euclidean Snyder model.

\section{The Snyder-Galilei harmonic oscillator in embedding coordinates}\label{sec:Embedding}

We can now embark into the investigation of the one-dimensional quantum harmonic oscillator in the framework of the Galilean Snyder model, starting from the simplest case which is given by the embedding coordinates.
First of all, we observe that
the phase space \coo $\hx_\mu$ and $\hpp_\mu$ can be realized in terms of canonical coordinates
$\bx_\mu$ and $\bp_\mu$ such that\footnote{From now on lovercase Greek indices run from $0$ to $1$, since we are working in $1+1$ dimensions.}.
 \begin{equation}
\left[\bar x_\mu,\bar x_\nu\right]= 0, \qquad \left[\bar x_\mu,\bar p_{\nu}\right]=i\hbar \eta_{\mu\nu}\qquad  \left[\bar p_{\mu},\bar p_{\nu}\right]=0 ,
\label{eq:canonical}
\end{equation}
the realization being given by:
\beq\lb{embed}
\hx_\mu=\sqrt{1-\be^2\bp_k^2}\ \bx_\mu,\qquad\hpp_\mu=\bp_\mu.
\eeq

Starting from this realization we can easily define a momentum  representation for the phase space operators, setting  $\bp_1\to p$, $\bx_1\to i\hbar{\deh\over\deh p}$, $\bp_0\to E$, $\bx_0\to-i\hbar{\deh\over\deh E}$, so that
\bea
&&\hat \pi_1=p,\qquad\hat x_1=i\hbar\embc\ {\deh\over\deh p},\lb{embs}\\
&&\hpp_0=E,\qquad\hx_0=-i\hbar\embc{\deh\over\deh E}\,,\lb{embt}
\eea
and the relevant Hilbert space is a (1+1)-dimensional space given by square integrable functions $\psi(p,E)$ of $p$ and $E$.

We assume that the  Hamiltonian for the  harmonic oscillator takes the standard form in terms of the Snyder spacetime coordinates and physical momenta:
\beq
H={\hpp_1^2\over2m}+{m\om^2\over2}\hx_1^2.
\eeq
In the \rep (\ref{embs})-(\ref{embt}) the associated \schr equation reduces to
\beq
E\ps={1\over2m}\left[-\hbar^2m^2\om^2\left((1-\be^2p^2){\deh^2\over\deh p^2}-\be^2p{\deh\over\deh p}\right)+p^2\right] \ps.
\eeq
The energy coordinate $E$ completely decouples from the equation, so one can set $\ps=\ps(p)$, with Hilbert space measure
\beq
d\Om={dp\over\embc},
\eeq
so that the space coordinate operator $\hx_1$ in (\ref{embs}) is Hermitian.

The \schr equation can be solved exactly in terms of Mathieu functions.
In fact, after defining a new variable $z=\be^\mo\arcsin\be p$, which induces a trivial measure in the Hilbert space, $d\Om=dz$, the \schr equation reads:
\beq\lb{mathieu}
{d^2\ps\over dz^2}-{1\over\al^2}\left({\sin^2\be z\over\be^2}-2mE\right)\ps=0,\qquad\qquad{\rm with}\quad\al=\hbar m\om.
\eeq

After another change of variables, defining $y=\be z$, eq.~(\ref{mathieu}) can be further simplified and written in the standard form of a Mathieu equation:
\beq
{d^2\ps\over dy^2}+(a-2q\cos2y)\ps=0,\qquad\qquad {\rm with} \quad q=-{1\over4\al^2\be^4},\quad a={-1+4\be^2mE\over2\al^2\be^4}. \label{mathieu2}
\eeq
The dimensionless variable $q$ so defined is  very large, and in this regime  the eigenvalues of (\ref{mathieu2}) can be obtained from eq.~(20.2.30) of \cite{AS}, giving 
\beq\lb{spectrum}
E_n=\hbar\om\left[\left(n+\ha\right)-{\al\be^2\over4}\left(n^2+n+\ha\right)+o(\al^2\be^4)\right].
\eeq

Alternatively, corrections to the standard energy spectrum for small $\beta$ can be found via a perturbative approach, writing
\beq
{\sin^2\be z\over\be^2}=z^2-{\be^2\over3}z^4+o(\be^4),
\eeq
so that the \schr equation (\ref{mathieu}) simplifies to
\beq
{d^2\ps\over dz^2}-{1\over\al^2}\left(z^2-{\be^2\over3}z^4-2mE\right)\ps=0.
\eeq
This is the equation of an anharmonic oscillator
and can be treated using standard perturbation theory. The solutions of the unperturbed equation are
\beq
\ps_n={1\over\sqrt{\sqrt{\pi\al}\,2^nn!}}\,e^{-z^2/2\al}\,H_n\left({z\over\sqrt\al}\right),\qquad\qquad E_n=\hbar\om\left(n+\ha\right),
\eeq
and the corrections to the spectrum to first order in $\beta^{2}$ are given by
\beq
\De E_n=-{\hbar\om\be^2\over3\al^2}\ (\ps_n,z^4\,\ps_n)=-{\hbar\om\al\be^2\over8}\left(n^2+n+\ha\right).
\eeq
Hence the energy levels of the harmonic oscillator read:
\beq
E_n=\hbar\om\left[\left(n+\ha\right)-{\al\be^2\over4}\left(n^2+n+\ha\right)+o(\al^2\be^4)\right],
\eeq
and one recovers (\ref{spectrum}).

\section{The Snyder-Galilei harmonic oscillator in Beltrami coordinates}\label{sec:Beltrami}
As we mentioned, we are interested in comparing the physical predictions of the Snyder model when  different choices for the physical momenta are taken. To this aim, in this section we derive the energy spectrum of the harmonic oscillator using a different set of momentum space coordinates than the one of the previous section, namely those associated to the Beltrami projective coordinates, defined in eq. (\ref{eq:Beltrami}).

Following the same steps as in the previous section, we start by identifying  a realization of the Galilean Snyder phase space (\ref{eq:GalileanSnyderBeltrami}) in terms of canonical coordinates (\ref{eq:canonical}):
\beq
\hx_0=\bx_0,\qquad\hx_i=\bx_i+\be^2\bp_i\bp_\al \bx_\al,\qquad\hp_\mu=\bp_\mu.
\eeq
In 1+1 dimensions this reduces to
\beq
\hx_0=\bx_0,\qquad\hx_1=(1+\be^2\bp_1^2)\bx_1-\be^2\bp_1\bp_0\bx_0,\qquad\hp_\mu=\bp_\mu,
\eeq
and allows us to find a representation on momentum space:
\beq\lb{beltrep}
\hat p_1=p,\qquad\hat x_1=i\hbar(1+\be^2p^2){\deh\over\deh p}+i\hbar\be^2p\,E{\deh\over\deh E},
\eeq
\beq
\hp_0=E,\qquad\hx_0=-i\hbar{\deh\over\deh E}.\lb{beltrep2}
\eeq
The Hilbert space of these operators is that of square integrable functions $\psi(p,E)$ of $p$ and $E$.
With the choice (\ref{beltrep}) of operator ordering, the measure in this Hilbert space is\footnote{It would also be possible to choose a \rep with
$\hx_0=-i\hbar\left({\deh\over\deh E}+\ha\right)$ and $d\Om={dp\,dE\over1+\be^2p^2}$, obtaining of course identical results.}
\beq
d\Om={dp\,dE\over(1+\be^2p^2)^{3/2}}.
\eeq

As before, we write the Hamiltonian for the  harmonic oscillator in the standard form in terms of the Snyder spacetime coordinates and physical momenta, which are now the Beltrami momenta:
\beq\lb{HamiltonianBeltrami}
H={\hp_1^2\over2m}+{m\om^2\over2}\hx_1^2.
\eeq

Using the representation (\ref{beltrep})-(\ref{beltrep2}), the \schr equation can be written as
\beq
\left[E-{p^2\over2m}\right]\ps=-{\al^2\over2m}\bigg[\ubp^2{\deh^2\over\deh p^2}+2\be^2\ubp p{\deh\over\deh p}\bigg(1+E{\deh\over\deh E}\bigg)
+\left(E^2{\deh^2\over\deh E^2}+2E{\deh\over\deh E}\right)\be^4p^2+\be^2E{\deh\over\deh E}\bigg]\ps, \label{eq:Schr}
\eeq
with $\al$ defined in (\ref{mathieu}).
Because on the r.h.s. of this equation the terms that contain $E$ are homogeneous, a possible ansatz for the eigenstate $\psi$ is
\beq
\ps(E,p)=E^\mu\ph(p).
\eeq
In this case, equation (\ref{eq:Schr}) reduces to an ordinary \schr equation for $\ph(p)$:
\beq
\left[\ubp^2{\deh^2\over\deh p^2}+2(1+\mu)\ubp\be^2p{\deh\over\deh p}+\left({2Em\over\al^2\be^2}+\mu\right)\be^2+\left(\mu^2+\mu-{1\over\al^2\be^4}\right)\be^4p^2\right]\ph=0.
\eeq

An equation of this form has been solved in \cite{Chang:2001kn} in the way we sketch in the following. Changing variables
\beq
z={p\over\sqrt{1+\be^2p^2}},\qquad\qquad{\rm with}\quad '={d\over dz},
\eeq
and defining
\beq
\ph=(\ubz)^{\la/2}f(\be z),
\eeq
one obtains
\beq
(\ubz)f''-(1-2\mu+2\la)\be zf'+\left(\mu-\la+{2Em\over\al^2\be^2}\right)f+\left(\mu^2+\mu+\la^2-\la-2\la\mu-{1\over\al^2\be^4}\right){\be^2z^2\over\ubz}f=0. \label{eq:Schr2}
\eeq
To have a normalizable solution the last term must vanish, fixing the value of $\la$ to
\beq\lb{mu}
\la=\mu+\ha\left(1\pm\sqrt{1+{4\over\al^2\be^4}}\right).
\eeq
Choosing the positive sign for the square root,  equation (\ref{eq:Schr2}) reduces to a Gegenbauer equation with eigenvalues
\beq\lb{spectrum2}
E_n=\hbar\om\left[\left(n+\ha\right)\left(\sqrt{1+{\al^2\be^4\over4}}+{\al\be^2\over2}\right)+{n^2\over2}\al\be^2\right],
\eeq
and the solution $\ps$  of the \schr equation (\ref{eq:Schr}) is:
\beq
\ps(E,p)={\rm const}\times E_n^\mu\,(\ubz)^{\la/2}C_n^\la(\be z).
\eeq
An important result is that in spite of the explicit dependence on the energy of the \schr equation, its eigenvalues only depend on the parameters $\al$ and $\be$. Thus, $E_n$ is independent of $\mu$, which can therefore be chosen arbitrarily. The most natural choice is $\mu=0$, namely
$\la=\ha\left(1+\sqrt{1+{4\over\al^2\be^4}}\right)$,
so that the solution $\ps$  of the \schr equation becomes independent of $E$ and can be normalized.

\section{The Snyder-Galilei harmonic oscillator in Poincar\'e coordinates}\label{sec:Poincare}

The final set of coordinates we will use in order to compare predictions of different choices for the physical momenta in the Snyder model  is that associated to the Poincar\'e projective coordinates, defined in eq. (\ref{eq:Poincare}). In 1+1 dimensions, \poi\coo lead to calculations rather similar to the previous case.

The realization of the Galilean Snyder phase space (\ref{eq:GalileanSnyderPoincare}) in terms of canonical coordinates (\ref{eq:canonical}) reads:
\beq
\hx_\mu=\left(1-{\be^2\over4}\bp_\al^2\right)\bx_\mu+{\be^2\over2}\bp_\mu\bp_\al\bx^\al,\qquad\tilde p_\mu=\bp_\mu.
\eeq
In 1+1 dimensions this reduces to
\beq
\hx_0=\bx_0,\qquad\hx_1=\left(1+{\be^2\over4}\bp_1^2\right)\bx_1-{\be^2\over2}\bp_1\bp_0\bx_0,\qquad\tilde p_\mu=\bp_\mu,
\eeq
and allows us to find the following representation on momentum space:
\beq\lb{poirep}
\tilde p_1=p,\qquad\hat x_1=i\hbar\left(1+{\be^2\over4}p^2\right){\deh\over\deh p}+i\hbar{\be^2\over2}p\,E{\deh\over\deh E},
\eeq
\beq
\tilde p_0=E,\qquad\hx_0=-i\hbar{\deh\over\deh E}.\lb{poirep2}
\eeq
The Hilbert space of these operators is that of square integrable functions $\psi(p,E)$ of $p$ and $E$.
With the choice (\ref{poirep}) of operator ordering, the measure in this Hilbert space is
\beq
d\Om={dp\,dE\over\ubq^{3/2}}.
\eeq

We again choose the standard Hamiltonian for the  harmonic oscillator, written now in terms of the Snyder-\poi coordinates:
\beq\lb{HamiltonianPoincare}
H={\tilde p_1^2\over2m}+{m\om^2\over2}\hx_1^2.
\eeq

Using the representation (\ref{poirep})-(\ref{poirep2}), the resulting \schr equation can be written as
\bea
\left(E-{p^2\over2m}\right)\ps=-{\al^2\over2m}\bigg[\ubq^2{\deh^2\over\deh p^2}+{\be^2\over2}\ubq p{\deh\over\deh p}\bigg(1+2E{\deh\over\deh E}\bigg)\nn\\
+\left(E^2{\deh^2\over\deh E^2}+{3\over2}E{\deh\over\deh E}\right){\be^4\over4}p^2+{\be^2\over2}E{\deh\over\deh E}\bigg]\ps,\label{eq:SchrPoincare}
\eea
with $\al$ defined in (\ref{mathieu}). Following the same steps as in the previous section, we make the ansatz
\beq
\ps(E,p)=E^\mu\ph(p),
\eeq
so that the \schr equation (\ref{eq:SchrPoincare}) reduces to
\beq
\left[\ubq^2{\deh^2\over\deh p^2}+\frac{1+2\mu}{2}\ubq\be^2p{\deh\over\deh p}+\left({2Em\over\al^2\be^2}+{\mu\over2}\right)\be^2+\left(\mu^2+{\mu\over2}
-{4\over\al^2\be^4}\right){\be^4\over4}p^2\right]\ph=0. \label{eq:SchrPoincare2}
\eeq
This can be solved as in the previous case by changing variables as
\beq
z={p\over2\sqrt{1+{\be^2\over4}p^2}},\qquad\qquad{\rm with}\quad '={d\over dz},
\eeq
and defining
\beq
\ph=\left(\ubz\right)^{\la/2}f(\be z).
\eeq
Then equation (\ref{eq:SchrPoincare2}) reads
\beq
(\ubz)f''-(1-4\mu+2\la)\be zf'+\left(2\mu-\la+{8Em\over\al^2\be^2}\right)f+\left(4\mu^2+2\mu+\la^2-\la-4\mu\la-{16\over\al^2\be^4}\right){\be^2z^2\over\ubz}\,f=0. \label{eq:SchrPoincare3}
\eeq
Again, the solution is normalizable if the last term vanishes, namely
\beq\lb{mup}
\la=2\mu+\ha\left(1\pm\sqrt{1+{64\over\al^2\be^4}}\right).
\eeq
Choosing the positive sign in (\ref{mup}),  equation (\ref{eq:SchrPoincare3}) reduces to a Gegenbauer equation with eigenvalues
\beq
E_n=\hbar\om\left[\left(n+\ha\right)\left(\sqrt{1+{\al^2\be^4\over64}}+{\al\be^2\over8}\right)+{n^2\over8}\al\be^2\right],
\eeq
and the eigenfunction of the \schr equation (\ref{eq:SchrPoincare}) is
\beq
\ps(E,p)={\rm const}\times E_n^\mu\,(\ubz)^{\la/2}C_n^\la(\be z).
\eeq
Also in this case $E_n$ is independent of $\mu$, and we can choose $\mu=0$, namely
$\la=\ha\left(1+\sqrt{1+{64\over\al^2\be^4}}\right)$,
so that  $\ps$  becomes independent of $E$ and normalizable.

\section{Conclusions}\label{sec:Conclusions}

The goal of this work was to investigate whether different choices of momentum space coordinates
 associated to the same noncommutative spacetime models are physically relevant. We did so by focussing on the well-known Snyder model, for which spacetime coordinates can be identified with the translation generators over a de Sitter manifold and momenta with coordinates on such manifold. We derived the energy spectrum of the harmonic oscillator in three different cases corresponding to the following choices of physical momenta, related to each other by a momentum space diffeomorphism: one where the physical momenta are those related to the embedding coordinates of the de Sitter manifold, one where they are related to the Beltrami projective coordinates and finally one where they are related to the Poincar\'e projective coordinates. We found that indeed these different choices of physical momenta imply different behaviours of the energy spectrum. In fact, the leading order contributions,  up to $\mathcal{O}(\beta^2)$, are:

\bigskip
\centerline{
\begin{tabular}{lc}
Embedding&$E_n=\hbar\om\left[n+\ha-{\hbar m \omega\be^2\over4}\left(n^2+n+\ha\right)\right]$\\[2ex]
Beltrami&$E_n=\hbar\om\left[n+\ha+{\hbar m \omega\be^2\over2}\left(n^2+n+\ha\right)\right]$\\[2ex]
\poi&$E_n=\hbar\om\left[n+\ha+{\hbar m \omega\be^2\over8}\left(n^2+n+\ha\right)\right]$\\
\end{tabular}}
\bigskip

In all cases, the leading corrections are qualitatively similar, but the coefficients of the correction terms beyond the standard contribution are different,  confirming that different parametrizations of the momentum space give rise to nonequivalent physical models.

Let us remark that this result was obtained by considering changes in the momentum space coordinates as active diffeomorphisms on the momentum space, to use the terminology of \cite{Amelino-Camelia:2019dfl, Amelino-Camelia:2016sru}.\footnote{Related to this,  notice that  diffeomorphisms in momentum space do not define "generalized canonical" transformations for the deformed phase space algebra, since they do not leave the algebra unchanged.} In fact, for each set $i$ of momentum space coordinates, $p_\alpha^{(i)}$,  we take the Hamiltonian of the harmonic oscillator to be written in the standard form in terms of these, namely $H={ \,\,p^{(i)}\,^2\over2m}+{m\om^2\over2}\hx_1^2$. The $p_\alpha^{(i)}$'s are thus identified with the physical momenta. Our results confirm, in a completely different setting and using a different model, the conclusions of \cite{Amelino-Camelia:2019dfl}, that active diffeomorphisms lead to different physical predictions. Of course, if we were to start from a Hamiltonian that takes a standard form in terms of some momentum space coordinates $p_\alpha^{(i)}$ and then simply mapped to some new momenta $p_\alpha^{(j)}$ by writing $p^{(i)}\left(p^{(j)}\right)$, we would be doing a mere change of variables, or a passive diffeomorphism, using the terminology of  \cite{Amelino-Camelia:2019dfl, Amelino-Camelia:2016sru}. In this case, the Hamiltonian  would take a different, non-standard, form in terms of the new momenta $p_\alpha^{(j)}$, $H={ \,\,p^{(i)}\left(p^{(j)}\right)\,^2\over2m}+{m\om^2\over2}\hx_1^2$,  but then we would not expect to see  physically relevant differences.

As a side result of our analysis, we were able to start comparing the predictions of the Galilean limit of the Snyder model, which we used in this work, to the predictions of the Euclidean Snyder model \cite{Mignemi:2011gr}, where only spatial coordinates are affected by noncommutativity. The latter had been widely  used to describe the modifications induced by Snyder-like spacetime noncommutativity on non-relativistic systems, see e.g. \cite{Mignemi:2011gr, Lu:2011it, Ivetic:2015cwa, Leiva:2012az, Quesne:2004pp}. In \cite{Ballesteros:2019mxi} it was shown that the proper non-relativistic limit of the Snyder model leads in fact to the Galilean Snyder model, which differs from the  Euclidean one because of a residual noncommutativity between the space and time coordinates. However, this difference might not be significant in the simple model we considered here, since  the energy levels of the harmonic oscillator in the Euclidean Snyder model \cite{Mignemi:2011gr} coincide with those obtained here with the Beltrami coordinates, eq.~(\ref{spectrum}), which are also the coordinates used in \cite{Mignemi:2011gr}. This can be traced back to the fact that, in the representation used here for the Beltrami coordinates, the spatial coordinate does depend on energy and its derivative, but in a  homogeneous way. So  the solutions to the \schr equation are separable and the part depending on momenta satisfies the same \schr equation as the solutions in the Euclidean model. One may wonder whether this is true also for the other choices of parametrization of the phase space considered in this paper or for less elementary models.  The first option does not seem to be the case. In fact, we notice that for  the  Poincar\'e coordinates we can make a similar remark to the one that applies to the Beltrami coordinates. Moreover, the representation of the embedding coordinates  does not mix spatial momenta and energy. We thus conjecture that the second option is more likely, since the only way to have  effects different from the Euclidean case would be to have a time-dependent Hamiltonian, so that the time coordinate introduces terms that are not homogeneous in the energy.

From a phenomenological point of view, one may estimate the size of the corrections to the quantum mechanical spectrum. These are of order $\hbar\om\be^2$.
As usual with the Snyder model, if one identifies $\be$ with the Planck length, the corrections are very small and hence not detectable experimentally.
For example, if one  considers an atomic-size oscillator with $m\om\sim10^{21}$kg/sec, they are of order $10^{-14}$.
However, one cannot exclude that the parameter $\be$ has a much greater value \cite{Sorkin:2007qi, DiCasola:2014xya} (see also discussion in \cite{Addazi:2021xuf} and references therein).

It would be interesting to further investigate the issues raised in the present paper, possibly considering other physical frameworks, especially in three spatial dimensions, such as the hydrogen atom, which could be compared with the results obtained in \cite{Ivetic:2015cwa} in the case of Euclidean Snyder model.

\section*{Acknowledgements}
The authors gratefully acknowledge  discussions with Angel Ballesteros during the early stages of this work. The authors would like to acknowledge the contribution of the COST Action CA18108. The authors acknowledge support from the INFN Iniziativa Specifica QUAGRAP.

\end{document}